*Article*

# Color Filtering Localization for Three-Dimensional Underwater Acoustic Sensor Networks

**Zhihua Liu [1], Han Gao [2], Wuling Wang [1], Shuai Chang [2] and Jiaxing Chen [2],***

[1]    College of Information Technology, Hebei Normal University, Shijiazhuang 050024, Hebei, China; E-Mail: hebtuliuzhihua@163.com; 19408272@qq.com.

[2]   College of Mathematics and Information Science, Hebei Normal University, Shijiazhuang 050024, Hebei, China; E-Mails: gaohan19920113@163.com; 981709087@qq.com.

*   Author to whom correspondence should be addressed; E-Mail:13930194955@163.com; Tel.: +86-0311-80786302; Fax: +86-0311-80786302.

External Editor:

**Abstract:** Accurate localization for mobile nodes has been an important and fundamental problem in underwater acoustic sensor networks (UASNs). The detection information returned from a mobile node is meaningful only if its location is known. In this paper, we propose two localization algorithms based on color filtering technology called PCFL and ACFL. PCFL and ACFL aim at collaboratively accomplishing accurate localization of underwater mobile nodes with minimum energy expenditure. They both adopt the overlapping signal region of task anchors which can communicate with the mobile node directly as the current sampling area. PCFL employs the projected distances between each of the task projections and the mobile node, while ACFL adopts the direct distance between each of the task anchors and the mobile node. Also the proportion factor of distance is proposed to weight the RGB values. By comparing the nearness degrees of the RGB sequences between the samples and the mobile node, samples can be filtered out. And the normalized nearness degrees are considered as the weighted standards to calculate coordinates of the mobile nodes. The simulation results show that the proposed methods have excellent localization performance and can timely localize the mobile node. The average localization error of PCFL can decline by about 30.4% than the AFLA method.

filtering; RGB (Red Green Blue); hierarchical structure

---

## 1. Introduction

Underwater acoustic sensor networks (UASNs) are composed of thousands of micro-sensors which are capable of sensing, operating, self-organizing and acoustic communicating to monitor underwater environment [1-3]. In UASNs, technology provides new capacity of marine resources exploration, pollution detection and aided navigation [4]. Nowadays, research of communication technology [3], architecture and protocol [3], localization and tracking algorithms [5], and network security [6] have been applied to various fields. As for a variety of applications in UASNs, the acquisition of information is meaningful only when the location is known for sensors. Underwater acoustic mobile localization (UABL) technology has been studied widely since it is a primary task used in topology control, coverage control and routing decision in UASNs [7, 8]. UABL typically employs a cluster of anchors whose locations could be obtained in advance and a set of mobile nodes whose locations are to be determined. Mobile nodes perform self-localization based on the information received from the anchors which belongs to the classification of distributed localization techniques [2]. The model of each mobile node equipped with a pressure sensor is motivated by reference [8] in which an anchor-free localization algorithm called AFLA was presented. AFLA was designed for active-restricted underwater sensor networks and made use of the relationship of adjacent nodes. There is a comprehensive survey of these UABL schemes in reference [2, 7]. As global positioning system (GPS) signals are highly weaken underwater, UABL algorithms often use rang-based methods to estimate distance, i.e., time of arrival (TOA) [9], angle of arrival (AOA) [10] and time difference of arrival (TDOA) [11].

In this article, a projection-color filtering localization algorithm called PCFL and an anchor-color filtering localization algorithm called ACFL are put forward. They both aim at cooperatively accomplishing precise localization for underwater mobile nodes with the minimum of power wastage. In the first place, the existing network construction is described as a hierarchical structure and the localization issue is converted into a geometry problem. Secondly, based on the task anchors which can communicate with the mobile node directly, task-rings are obtained considering the task projections (i.e. projections of the task anchors) as centers, and samples are randomly selected in the overlapping area of the task-rings. Later, the RGB sequences for both the mobile nodes and the samples are computed based on the projection distances. Different from the existing CDL algorithms utilizing the DV-hop measurement [12], PCFL uses the AOA measurement and the initial RGB values are given to the task projections. While for ACFL, the initial RGB values are given to the task anchors. Last, the nearness degree is defined to filter samples, and at the same time nearness degrees are stored as weights.

The rest of this paper is arranged as follows. In Section 2, we survey some of the existing range-free localization techniques in UASNs. In Section 3, PCFL and ACFL algorithms are put forward and the performance is verified, where we deduce the task-rings sampling method, compute the nearness degree threshold for filtering samples and locate the mobile nodes weightly. Simulations performance of PCFL and ACFL is evaluated in Section 4, and in the end conclusions are given in Section 5.

## 2. Background

### 2.1. Related studies

Some studies have researched the range-free algorithms, these range-free algorithms do not need to estimate the distance precisely [13-15]. Instead, sensors can localize themselves by taking advantage of signal connectivity, delay time, hop-distance and angle information. Schemes are always more economic and simpler than the range-based ones, but less accurate.

CDL (color-theory based dynamic localization) [12], FRORF [14] and DV-hop algorithms [13] are range-free methods which did not acquire distance information. But they always have coarser performance. CDL calculated RGB (Red, Green, Blue) sequences based on DV-hop for both samples and mobile nodes. It converted RGB to HSV (Hue, Saturation, Value) for mobile nodes and samples using the traditional convert algorithm. The color theory indicated that the information of RGB and HSV fused different data of red、green and blue. According to the color theory, only V (value) of HSV changed in proportion to the distances between anchors and the mobile nodes. After the new HSV was obtained, the another RGB data of the nodes was acquired by the HSV to RGB method. Then CDL filtered the nearest sample by searching for the most similar RGB sequence to the mobile node, and identified it as the location of the mobile node. Although CDL has better location performance in various terrestrial networks, it's not suitable for localization in UASNs as the inconsistency distribution of nodes and the weak association between signal hops and actual distances. Liu et al. proposed a local sampling and filtering color dynamic localization (LSF-CDL) [16]. Using the collected signals, LSF-CDL adopted the overlapping signal region of anchors which were able to communicate with the mobile node directly as the new local sampling area. Also the proportion factor of distance was used to weight the average hop distance which optimized the calculation of hop distance in CDL. By comparing the RGB difference sequences, samples could be filtered out. FRORF method represented overlapping rings as fuzzy sets to isolate a region where the node was most likely located [14]. And the localization accuracy was improved under different number of anchors and degrees of radio propagation irregularity.

Compared to terrestrial nodes which can keep their location unchanged after arrangement, underwater nodes are influenced by tide、ocean current and other factors leading to their locations unfixed [1]. Due to the influence of node mobility、multipath fading and shadow、long time delay、the variation of sound velocity and asymmetry factor, there are more challenges for the UABL algorithms [17]. The AUV-aid (autonomous underwater vehicle aid) method employed a large number of sensors and one AUV for balancing performance and cost. AUV was used for localization and carrying messages of disconnected sensors or time-critical information [18]. The Anchor-based method relied on the TDOA locally measured at a sensor to detect range differences from the sensor to four anchors [19]. The AFLA method was a self-localization algorithm designed for anchor-free UASNs [8]. However, these schemes have not provided sufficient accuracy. In order to improve localization performance, we propose two new algorithms: PCFL and ACFL here, and aim at projecting the locations of task anchors to the mobile node's plane which converts the 3D UABL problem to 2D.

### 2.2. Network model

There is a typical UASNs model as shown in Figure 1. There are three types of nodes in the model: mobile nodes, anchors and GPS devices. GPS devices are drifting on the water surface and are often equipped with GPS to get their absolute locations or by other means. Anchors are vertical hanging below the GPS devices in order to obtain the two-dimensional coordinates. The main role of the anchors is helping the mobile nodes to finish self-localization.

**Figure 1.** Underwater acoustic sensor networks structure

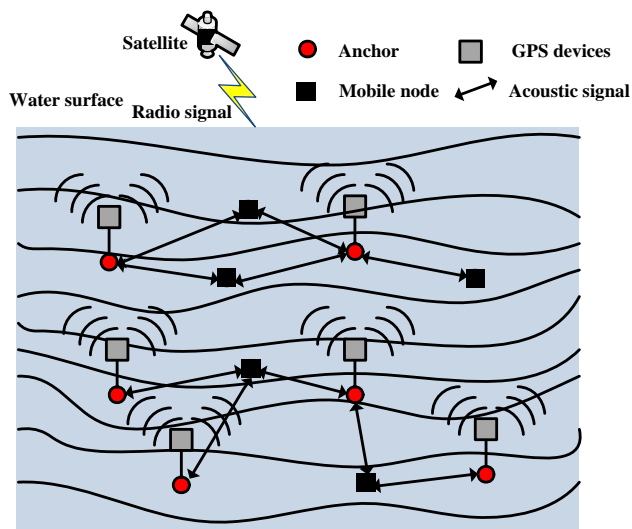

### 2.3. Problem formulation

With the purpose of developing an accurate and high-performance UABL used in the complex underwater environments, the assumptions are made as below:

Both the anchors and the mobile nodes have the same communication range $R$ and are equipped with a few array wireless antennas in order to communicate with each other by acoustic signals which are used to transmit the depth information, the anchors' coordinates and the measured AOA values.

The mobile node transmits acoustic signal regularly. Once receiving the acoustic signal, each anchor replies a message including its own three dimensional coordinates and its depth information.

When the information exchange is completed, the mobile node has sufficient information to localize itself. The information refers to locations and the depth information of all task anchors and the AOA measurement from the task anchors estimated by the mobile node.

The depth information is measured by the pressure sensors which are equipped within all the nodes.

In UASNs, it is difficult to achieve the time synchronization precisely as a result of the characteristics of acoustic signals propagation. While by employing the AOA values of the acoustic signals from the anchors, synchronous request between the nodes is not so necessary. So, PCFL and ACFL techniques, advanced here, have the advantage on synchronization.

Let's consider a 3D UASNs $UN$ with $p$ anchors and $q$ mobile nodes and express the location of each node as

$$n_i = (x_i, y_i, z_i), n_i \in UN, i = 1, 2, \ldots p + q \tag{1}$$

We assume that the locations of $p$ anchors in $A = \{n_1, n_2, \ldots n_p\}$ are known but the locations of the other $q$ mobile nodes in $M = \{n_{p+1}, n_{p+2}, \ldots n_{p+q}\}$ are unknown and to be localized, $UN = A \cup M$. In our localization methods, the only obtainable information is the smaller angle of acoustic signal $\alpha_{ij}$ from the anchors $n_j$ ($j = 1, 2, \ldots p$) received by the mobile node $n_i$ ($i = p + 1, \ldots p + q$) based on the AOA measurement and the depth information obtained by the corresponding pressure sensors. The Euclidean distance between the node $n_i$ and the node $n_j$ can be calculated as

$$d(n_i, n_j) = \sqrt{(x_i - x_j)^2 + (y_i - y_j)^2 + (z_i - z_j)^2}, (n_i \in UN, n_j \in UN) \tag{2}$$

At time instant $t$, the set of task anchors corresponding to the mobile node $n_i$ can be expressed as

$$A_i^t = \{n_j \,\big|\, d(n_i, n_j) \leq R, i = p + 1 \ldots p + q, j = 1, 2, \ldots p\} \subset A \tag{3}$$

Let $p^t$ denote the number of task anchors in $A_i^t$, it can be seen that

$$\left| A_i^t \right| = p^t \leq p \tag{4}$$

As all the nodes have the same communication range $R$, hence, at time instant $t$, the mobile node $n_i$ ($i = p + 1, \ldots p + q$) can exchange information with anchor $n_j$ directly if and only if $n_j \in A_i^t$. The set of task projections $B_i^t$ for $n_j \in A_i^t$ is defined as

$$B_i^t = \{n_j' \,\big|\, d(n_i, n_j) \leq R \wedge x_j' = x_j \wedge y_j' = y_j \wedge z_j' = z_i\}, \left| B_i^t \right| = p^t \leq p \tag{5}$$

Where $n_j' = (x_j', y_j', z_j')$. RGB sequences for the projections in $B_i^t$ and RGB sequences for the anchors in $A_i^t$ at time instant $t$ are randomly assigned numerical values in the range of [0,1] as stated in the reference [12] and respectively are defined as

$$RGB_j^{pt} : \{R_j^{pt}, G_j^{pt}, B_j^{pt}\}, (n_j' \in B_i^t) \text{ and } RGB_j^{at} : \{R_j^{at}, G_j^{at}, B_j^{at}\}, (n_j \in A_i^t) \tag{6}$$

The task-ring for task projection $n_j' \in B_i^t$ is defined as

$$C_j^t = \{(x, y, z) \,\big|\, (x - x_j')^2 + (y - y_j')^2 \leq (R^2 - (k_{ij}^t)^2), z = z_j'\} \tag{7}$$

Where $k_{ij}^t$ ($j = 1, \ldots p^t, i = p + 1, \ldots q$) is the depth difference between the task anchors $n_j$ and the mobile node $n_i$ which can be calculated using the information from their deployed pressure sensors. Note that the minimum communication angle between the task anchors $n_j$ and the mobile node $n_i$ at time instant $t$ named as $\alpha_{ij}^t$ ($j = 1, \ldots p^t, i = p + 1, \ldots q$) can be measured by the mobile node $n_i$, then the localization issue can be described as follows:

$$Estimate \quad n_i$$
$$subject \quad to \quad n_j(n_j \in A_i^t)(n_j'(n_j' \in B_i^t)); \alpha_{ij}^t; k_{ij}^t; (i = p+1, ...q) \tag{8}$$

## 3. Algorithm design

In this section, we present our color filtering localization algorithms, which we call projection-color filtering localization (PCFL) and anchor-color filtering localization (ACFL).

Firstly, we put forward the three dimensional hierarchical structure according to the depth of the anchors and the mobile nodes. And then we introduce the design details of PCFL and ACFL methods. In the end, we show the feasibility's analysis of them. PCFL and ACFL are both based on the color theory. The distances between nodes are calculated by the projection method, while CDL calculates the distance based on DV-hop method.

### 3.1. Hierarchical structure model

In UASNs, the most important characteristics are node mobility, multipath propagation loss, time uncertainty, and low communication rate. In this paper, our two localization methods are based on hierarchical structure model.

Figure 2 shows a hierarchical structure and projection model. Using the depth information of the mobile node, three task anchors $n_1, n_2$ and $n_3$ will be projected to three positions $n_1', n_2'$ and $n_3'$ in the mobile node plane, respectively.

**Figure 2.** Hierarchical and projection structure model

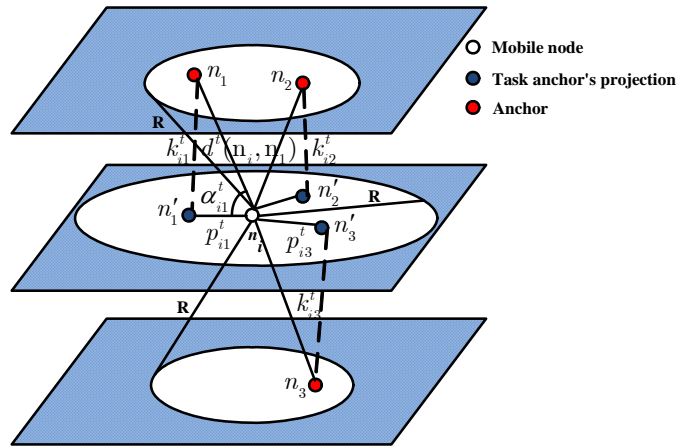

Note that the minimum communication angle $\alpha_{ij}^t \in [0, \frac{\pi}{2}]$ ($j = 1, 2, 3$) can be measured by the mobile node $n_i$ based on the AOA measurement, and the depth difference $k_{ij}^t$ ($j = 1, 2, 3$) can be measured by the deployed pressure sensors. When the anchor $n_j$ and the mobile node $n_i$ are with the same depth, $\alpha_{ij}^t = 0$, $k_{ij}^t = 0$. When $\alpha_{ij}^t = \frac{\pi}{2}$, the value of the minimum communication angle between $n_j$ and $n_i$ will be maximum. So at time instant $t$ the geographic distance between the anchor $n_j$ and the mobile node $n_i$ can be calculated as

$$d^t(n_i, n_j) = k_{ij}^t / \sin \alpha_{ij}^t \qquad (9)$$

And at time instant $t$, the distance $p_{ij}^t$ between the mobile node $n_i$ and the task anchor's projection $n_j'$ can be calculated as

$$p_{ij}^t = k_{ij}^t / \tan \alpha_{ij}^t \qquad (10)$$

### 3.2. PCFL and ACFL

The PCFL algorithm and the ACFL algorithm can be both divided into three steps: (a) determination of the sampling area; (b) RGB values calculation, and (c) filtering and weighted evaluation.

(a) In the first step, we use the task-rings sampling method to determine the sampling area. Here every task-ring is expressed as the signal range interface for every corresponding task anchor, as defined in formula (7) previously. Then, the sampling area constraining the location of the mobile node is the intersection region of these task-rings corresponding to each task projection.

For instance, the shadow region in Figure 3 represents the sampling area which is derived from the task-rings of task projections $n_1'$, $n_2'$ and $n_3'$ with respect to task anchors $n_1$, $n_2$ and $n_3$, respectively. The task-rings are represented as the blue dotted circles. As each task anchor can receive the signal from the mobile node, it can be inferred that the mobile node's location is inside its task-ring. And the following theorem 1 will continue to prove this. The following restricted conditions based on the task-rings sampling method will confine the samples shown as the purple squares inside the shadow region in Figure 3:

$$(x - x_1')^2 + (y - y_1')^2 \le (R^2 - (k_{i1}^t)^2) \wedge z = z_1' = z_i$$
$$(x - x_2')^2 + (y - y_2')^2 \le (R^2 - (k_{i2}^t)^2) \wedge z = z_2' = z_i$$
$$(x - x_3')^2 + (y - y_3')^2 \le (R^2 - (k_{i3}^t)^2) \wedge z = z_3' = z_i \qquad (11)$$

Where $(x, y, z)$ denotes the coordinates of the sample. Normally, the set of the sampling area $\mathrm{S}_i^t$ is defined as

$$\mathrm{S}_i^t = \{(x, y, z) | \forall n_j' \in \mathrm{B}_i^t, (x - x_j')^2 + (y - y_j')^2 \le (R^2 - (k_{ij}^t)^2) \wedge z = z_i\} \qquad (12)$$

(b) The process of RGB values calculation can be divided into two stages. The first stage is the RGB sequences calculation for the mobile nodes, denoted by $RGB_i^{Mt} : \{R_i^{Mt}, G_i^{Mt}, B_i^{Mt}\}, (n_i \in M)$. After arranging all the RGB values $RGB_j^{pt} : \{R_j^{pt}, G_j^{pt}, B_j^{pt}\}, (n_j' \in B_i^t)$ for task anchors' projections (PCFL) or arranging all RGB values $RGB_j^{at} : \{R_j^{at}, G_j^{at}, B_j^{at}\}, (n_j \in A_i^t)$ for task anchors (ACFL), the mobile node converts $RGB_j^{pt}$ or $RGB_j^{at}$ into HSV (Hue, Saturation, and Value) [12] at time instant $t$.

$$H_j^{kt} S_j^{kt} V_j^{kt} = (RGB \text{ to } HSV) R_j^{kt} G_j^{kt} B_j^{kt}; k = p, a \qquad (13)$$

**Figure 3.** The sampling area

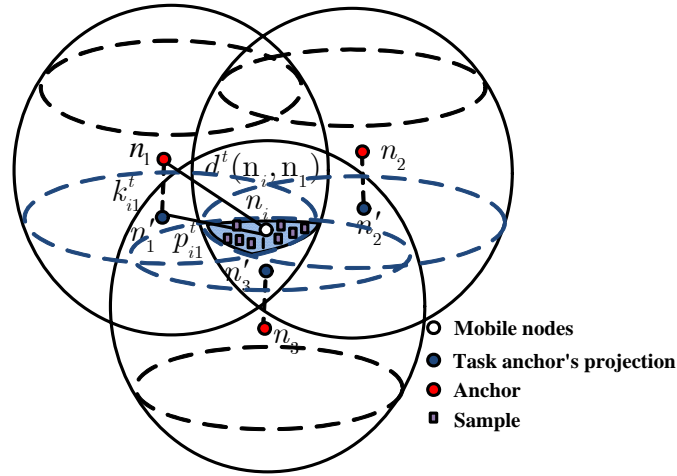

And then we can calculate the changed values $H_{ij}^{kt} \ S_{ij}^{kt} \ V_{ij}^{kt}$ of HSV between $n_i$ and the task projection $n_j'$ (or task anchor $n_j$) by the equations below

$$H_{ij}^{kt} = H_j^{kt}, S_{ij}^{kt} = S_j^{kt}, V_{ij}^{kt} = (1 - \frac{d_{ij}^t}{Range}) \times V_j^{kt}; k = p, a; d_{ij}^t = p_{ij}^t, \text{or} \quad d(n_i, n_j) \qquad (14)$$

Where *Range* is assumed to be the maximum length of color change, here we assume *Range* to be the maximum communication distance of the mobile node's plane. Then $\{R_{ij}^{kt}, G_{ij}^{kt}, B_{ij}^{kt}\}$ between the mobile node $n_i$ and the task projection $n_j'$ (or task anchor $n_j$) can be worked out based on the algorithm *HSV to RGB* [12].

$$R_{ij}^{kt} G_{ij}^{kt} B_{ij}^{kt} = (HSVtoRGB) H_{ij}^{kt} S_{ij}^{kt} V_{ij}^{kt}; k = p, a \qquad (15)$$

At time instant *t*, the mobile node $n_i$ records the task anchors in $A_i^t$ and their depth information, then computes the distances from $n_i$ to the projections in $B_i^t$ for PCFL (or computes the distances from $n_i$ to the anchors in $A_i^t$ for ACFL), and $n_i$ processes normalization of these distances to calculate the proportion factor of distances weights $\lambda_{ij}^t$ defined as below:

$$\lambda_{ij}^t = \frac{(d_{ij}^t)^{-1}}{\sum\limits_{n_j \in A_i^t} (d_{ij}^t)^{-1}}, d_{ij}^t = p_{ij}^t, \text{or} \ d^t(n_i, n_j) \qquad (16)$$

At last using the weighted mean of $\{R_{ij}^{kt}, G_{ij}^{kt}, B_{ij}^{kt}\}$, RGB values $\{R_i^{Mt}, G_i^{Mt}, B_i^{Mt}\}$ for the mobile node $n_i$ can be worked out.

$$\{R_i^{Mt}, G_i^{Mt}, B_i^{Mt}\} = \sum\limits_{n_j \in A_i^t} \lambda_{ij}^t \{R_{ij}^{kt}, G_{ij}^{kt}, B_{ij}^{kt}\}; k = p, a \qquad (17)$$

The second stage is the RGB sequences calculation for samples, the RGB sequence for the sample $s_k$ is denoted by $RGB_i^{s_k t} : \{R_i^{s_k t}, G_i^{s_k t}, B_i^{s_k t}\}$. The calculation method is similar to the above, and the difference just lines in replacing $d_{ij}^t$ in the above formula (16) with the Euclidean distance between the mobile node $n_i$ and the sample $s_k$ which can be calculated using formula (2).

(c) filtering and weighted evaluation

PCFL algorithm and ACFL algorithm both filter the samples in the sampling area based on nearness degree. Assume at time instant $t$, PCFL and ACFL both sample $m$ times randomly and the RGB sequence of sample $s_k$ is $\{R_i^{s_k t}, G_i^{s_k t}, B_i^{s_k t}\}$, then the nearness degree $\mu_{s_k M}^t$ between the mobile node $n_i$ and sample $s_k$ is defined as

$$\mu_{s_k M}^t = \sqrt{(R_i^{s_k t} - R_i^{Mt})^2 + (G_i^{s_k t} - G_i^{Mt})^2 + (B_i^{s_k t} - B_i^{Mt})^2} \ \ (i = p+1,...q; k = 1, 2,...m) \quad (18)$$

Then the filtered samples set $\tilde{S}_i^t$ is

$$\tilde{S}_i^t = \{s_k \mid s_k \in S_i^t \wedge \mu_{s_k M}^t \le \mu^t\}, n_i \in M \quad (19)$$

Where $\mu^t$ is the threshold at time instant $t$. The relationship between $\mu^t$ and the localization error can be seen at the simulation part in Figure 5. Assume there are $m^t (m^t \le m)$ samples have been filtered out, from formulas (18) and (19), the following theorem 2 will prove the fact that the closer the mobile node $n_i$ is to the sample $s_k$, the smaller the nearness degree $\mu_{s_k M}^t$ is. Therefore, based on $\mu_{s_k M}^t$, PCFL and ACFL do the normalized weighted processing in the calculation of $n_i$ coordinates.

$$\tilde{\mu}_{s_k M}^t = \mu_{s_k M}^t (\sum \mu_{s_k M}^t)^{-1}, s_k \in \tilde{S}_i^t \quad (20)$$

Assume the coordinate of the filtered sample $s_k$ is $(x_{s_k}^t, y_{s_k}^t, z_i^t)$, then the coordinate $(x_i^t, y_i^t, z_i^t)$ of the mobile node $n_i$ at time instant $t$ can be calculated using $\tilde{\mu}_{s_k M}^t$ as the weights, note that $z_i^t$ can be achieved by the deployed pressure sensor.

$$x_i^t = \sum_{s_k \in \tilde{S}_i^t} x_{s_k}^t \tilde{\mu}_{s_k M}^t, y_i^t = \sum_{s_k \in \tilde{S}_i^t} y_{s_k}^t \tilde{\mu}_{s_k M}^t \quad (21)$$

Figure 4 gives the architecture for both the PCFL algorithm and the ACFL algorithm.

*3.3. Feasibility's analysis*

In this section, we describe the feasibility of the PCFL algorithm and the ACFL algorithm. The following theorems are to prove the rationality of them.

Theorem 1. Assume that $n_i$ is the mobile node, and $n_j'$ is the task projection of task anchors $n_j (n_j \in A_i^t)$ for $n_i$, that is, $d(n_j, n_i) \le R$ and $x_j = x_j', y_j = y_j'$. Then, $n_i$ is inside the intersection area of the rings taking $n'_j$ as the center and $\sqrt{R^2 - (k^t_{ij})^2}$ as the radius, correspondingly.

**Figure 4.** The architecture for PCFL and ACFL algorithm

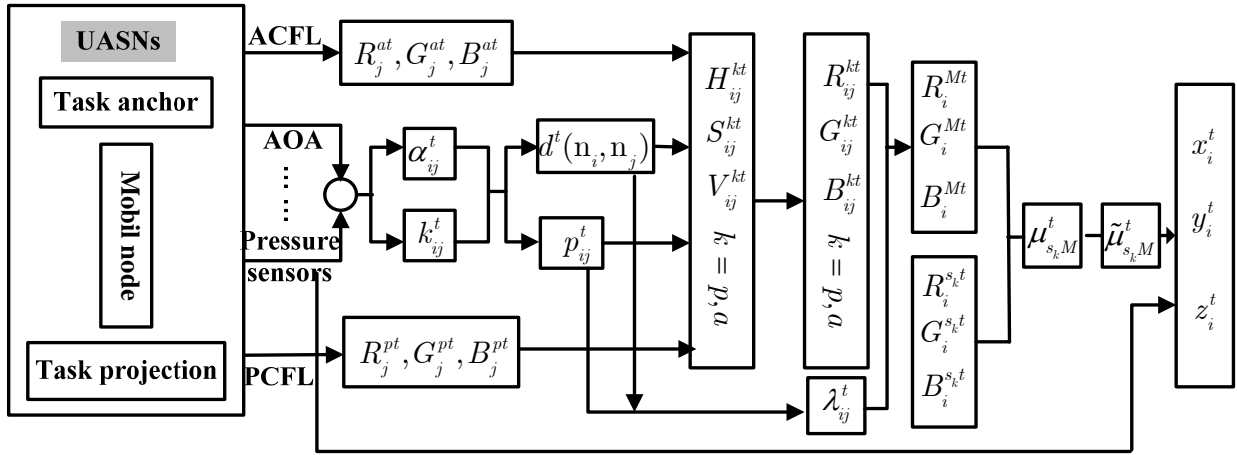

Proof. As for all $n_j \in A^t_i$, there is $d(n_j, n_i) \le R$, we can see that $n_i$ is in the intersection area of the spheres taking $n_j$ as the centers and $R$ as the radius, respectively, as shown in Figure 3. While, every sphere intersects the $n_i$ plane so as to get its ring. In view of these rings inside the $n_i$ plane would be the task-rings defined in formula (7). Then, $n_i$ is in the intersection of these task-rings, such as one of them takes $n'_j$ as the center and $\sqrt{R^2 - (k^t_{ij})^2}$ as the radius. The theorem can also verify correctness of the constraints in formula (11) and formula (12) for the sampling area. □

Lemma 1. Each RGB sequence is unique.

Proof. The lemma can be proved by *redectio ad absurbum*. Suppose that in PCFL algorithm and ACFL algorithm at time instant $t$ there are two different mobile nodes (or samples) with the same RGB sequence. This indicates that the two sets of corresponding task anchors are the same, and the two corresponding distances or corresponding projection distances are also same, although the mobile nodes (or samples) are different. However, this is a contradiction as, by the trilateration algorithm, two different nodes (or samples) in a three-dimensional region are the same when the distances between four anchors are known. □

Theorem 2. The smaller the nearness degree $\mu^t_{s_k M}$ is, the closer sample $s_k$ gets to mobile node $n_i$.

Proof. It is clear that $\{R^{Mt}_i, G^{Mt}_i, B^{Mt}_i\}$ is unique to the mobile node $n_i$ and $\{R^{s_k t}_i, G^{s_k t}_i, B^{s_k t}_i\}$ is unique to the sample $s_k$ from Lemma 1. Consider two samples, respected as $s_l$ and $s_k$, respectively. Next, we assume that $\tilde{d}^t_{lj}$ ($\tilde{d}^t_{kj}$) denotes the distance between the sample $s_l$ ($s_k$) and the task projection $n'_j$ (or task anchor $n_j$), respectively. And assume that

$$\mu^t_{s_l M} \le \mu^t_{s_k M} \qquad (22)$$

Then, the theorem can be proved by *redectio ad absurbum*. Suppose the sample $s_k$ is closer to the mobile node $n_i$ than the sample $s_l$ and

$$\left| \tilde{d}_{lj}^t - d_{ij}^t \right| > \left| \tilde{d}_{kj}^t - d_{ij}^t \right|, (d_{ij}^t = p_{ij}^t \text{ or } d^t(n_i, n_j)) \tag{23}$$

So

$$\left| \tilde{V}_{lj}^{ft} - V_{ij}^{ft} \right| > \left| \tilde{V}_{kj}^{ft} - V_{ij}^{ft} \right|; f = p, a \tag{24}$$

Where $\tilde{V}_{lj}^{ft}(\tilde{V}_{kj}^{ft})$ denotes the V value of HSV between the sample $s_l(s_k)$ and the task projection $n_j'$ (or task anchor $n_j$). Then, by formula (16),

$$\left| \tilde{\lambda}_{lj}^t - \lambda_{ij}^t \right| > \left| \tilde{\lambda}_{kj}^t - \lambda_{ij}^t \right| \tag{25}$$

And by formula (13) to formula (15), we can see

$$\left| \tilde{R}_{lj}^{ft} - R_{ij}^{ft} \right| = \left| \tilde{R}_{kj}^{ft} - R_{ij}^{ft} \right|, \left| \tilde{G}_{lj}^{ft} - G_{ij}^{ft} \right| = \left| \tilde{G}_{kj}^{ft} - G_{ij}^{ft} \right|, \left| \tilde{B}_{lj}^{ft} - B_{ij}^{ft} \right| > \left| \tilde{B}_{kj}^{ft} - B_{ij}^{ft} \right| f = p, a \tag{26}$$

So by formula (17)，we have

$$\left| R_i^{s_l t} - R_i^{Mt} \right| > \left| R_i^{s_k t} - R_i^{Mt} \right|, \left| G_i^{s_l t} - G_i^{Mt} \right| > \left| G_i^{s_k t} - G_i^{Mt} \right|, \left| B_i^{s_l t} - B_i^{Mt} \right| > \left| B_i^{s_k t} - B_i^{Mt} \right| \tag{27}$$

Then

$$\mu_{s_l M}^t > \mu_{s_k M}^t \tag{28}$$

However, this is a contradiction to the formula (22).      □

Next we will analysis the complexity of the worst case time and the worst case space for PCFL and ACFL.

Theorem 3. The new algorithms take $\mathrm{O}(n)$ worst case time and $\mathrm{O}(n)$ worst case space.

Proof. Let's consider a 3D UASNs $UN$ with $p$ anchors as stated before, in the process (a) determination of the sampling area, it takes $\mathrm{O}(p)$ time for the mobile node to obtain the angles respected to each task anchor using the AOA method and it requires $\mathrm{O}(p)$ space to store angles. Then it takes $\mathrm{O}(p)$ time to compute the distances from the mobile node to every task anchor according to the formula (9) or to every task projection according to the formula (10) and it requires $\mathrm{O}(p)$ space to store the depth information and $\mathrm{O}(p)$ space to store the distance. At last it takes $\mathrm{O}(p)$ time to confine the samples and it requires $\mathrm{O}(p)$ space to store these samples.

In the process (b) RGB values calculation, it first takes $\mathrm{O}(p)$ time for the mobile node to convert the RGB values of the task anchors or of the task projections into HSV according to the formula (13) and it requires $\mathrm{O}(p)$ space to store these HSV values. Then it takes $\mathrm{O}(p)$ time to calculate the changed HSV values according to the formula (14) and it requires $\mathrm{O}(p)$ space to store the changed HSV values. Later it takes $\mathrm{O}(p)$ time to work out the changed RGB values according to the formula (15) and it requires $\mathrm{O}(p)$ space to store the changed RGB values. At last it takes $\mathrm{O}(p)$ time to work out the proportion factor of distances weights and the RGB values for the mobile node according to the formula (16) and formula (17) and it requires $\mathrm{O}(p)$ space to store the proportion factor of distances weights and the RGB values for the mobile node.

In the process (c) filtering and weighted evaluation, assuming sample $m$ times randomly, then it takes $\mathrm{O}(m)$ time to work out the nearness degree according to the formula (18) and it requires $\mathrm{O}(m)$ space to store the nearness degree. It takes $\mathrm{O}(m)$ time to filter out the samples according to the formula (19) and it requires $\mathrm{O}(m)$ space to store the filtered samples. At last it takes $\mathrm{O}(m)$ time to normalize weighted the nearness degree and calculate the coordinates of the mobile node according to the formula (20) and the formula (21) and it requires $\mathrm{O}(m)$ space to store the normalized weighted nearness degree and the coordinates of the mobile node.

Therefore, in total, both PCFL and ACFL can be considered as taking $\mathrm{O}(n)$ time and $\mathrm{O}(n)$ space in the worst case to estimate the coordinates of the mobile node.                                    □

Table 1 lists the comparison of worst case computational complexities between the typical algorithms and our new algorithms.

**Table 1.** Comparison of worst case time

|       | PCFL/ACFL | Anchor-aid | AUV-aid | AFLA |
|-------|-----------|------------|---------|------|
| Time  | $\mathrm{O}(n)$ | $\mathrm{O}(n \log n)$ | $\mathrm{O}(n \log n)$ | $\mathrm{O}(n^2)$ |
| Space | $\mathrm{O}(n)$ | $\mathrm{O}(n)$ | $\mathrm{O}(n)$ | $\mathrm{O}(n)$ |

## 4. Simulation results

In this section, we make a comprehensive evaluation for the PCFL algorithm and the ACFL algorithm through simulation experiments with Matlab 7.0. Three localization schemes, Anchor-based [19]、AUV-aid [18] and AFLA [8] are compared with PCFL and ACFL. In the simulation experiments, the localization area is defined as $1000m \times 1000m \times 20m$ where anchors are randomly deployed. The depth information of mobile nodes and anchors can be obtained by the pressure sensors at each time instant $t$. The original speed of mobile nodes which move with ocean water is assumed to be 0.1m/s. The maximum communication radius $R$ is assumed to be 100m. The localization error $E_i^t$ for the mobile node $n_i, (i = p+1, \ldots p+q), \; n_i \in M$ at time instant $t$ can be calculated by [13]:

$$E_i^t = \sum_{k=1}^{T} \sqrt{(\widehat{x}_i^t - x_i^t) + (\widehat{y}_i^t - y_i^t) + (\widehat{z}_i^t - z_i^t)} \Big/ T \tag{29}$$

$(\widehat{x}_i^t, \widehat{y}_i^t, \widehat{z}_i^t)$ and $(x_i^t, y_i^t, z_i^t)$ are the real coordinate and the estimated coordinate of the mobile node $n_i$, respectively. We run the simulation 50 times for each group of data, namely $T = 50$.

### 4.1. The localization error of ACFL and PCFL under different parameters

We compare localization performance of PCFL and ACFL through three different variables which are the threshold, the density of anchors and the number of samples.

**The threshold ($\mu^t$)**: Figure 5 shows the localization error for both ACFL and PCFL changing with the threshold $\mu^t$. The density of anchors is set for 4 and the number of samples is set for 400 in Figure 5. In general, the larger the threshold, the less number the task anchors which can communicate with mobile nodes. The localization error of ACFL (PCFL) reaches the minimum, when the threshold is set for 0.0142 and 0.01, respectively. The localization error is increasing as the growth of the threshold. In general, the average localization error of ACFL is bigger than that of PCFL as shown in Figure 5. When the threshold is changed from 0.01 to 0.03, the average localization error of PCFL is less than 2m, while that of ACFL is more than 4m.

**The number of samples**: Filtering out more samples for ACFL (PCFL) in the process of simulation experiments can improve localization error, but needing more energy consumption. Considering this situation, the number of samples which are deployed to localize mobile nodes should be reasonable for a compromise between energy saving and localization accuracy. In Figure 6 the threshold is set for 0.0142 and the density of anchors is set for 4. When the number of the deployed samples which are filtered out to localize mobile nodes is much enough, samples are concentrated in the vicinity of the mobile node and the localization error reduces obviously. We indicate that the localization error for both PCFL and ACFL is 0.91m and 4.44m when the number of the deployed samples is 500, respectively.

| **Figure 5.** The threshold | **Figure 6.** The number of samples |
|---|---|
| 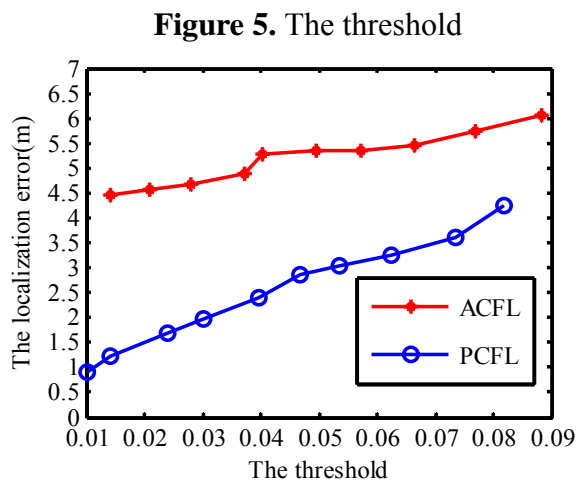 | 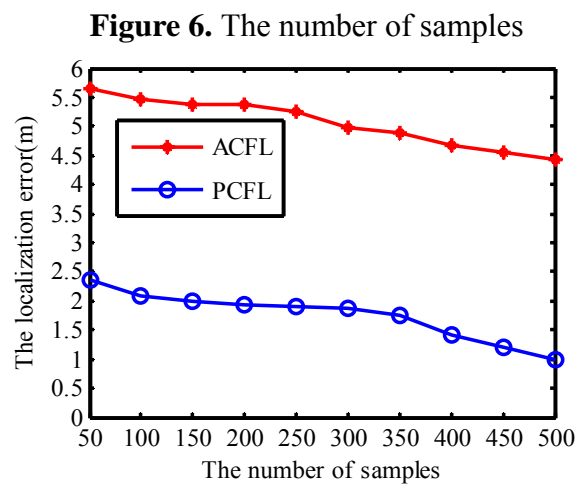 |

**The density of anchors**: In Figure 7 the threshold is set for 0.0142 and the number of samples is set for 400. $N_{anchor}$ stands for the average number of task anchors in the communication radius $R$ and is assumed to change from 5 to 100. We assume that $V$ denotes the volume of the simulation space. $D_{anchor}$ is defined as the density of task anchors and it can be calculated by the formula $D_{anchor} = (\frac{4}{3}\pi R^3)N_{anchor} / V$ [16], so we can obtain different values of the localization error when $D_{anchor}$ varies from 0.5 to 5 through the simulation experiment. We can filter out more task anchors with the increase of $D_{anchor}$. The localization error of PCFL (or ACFL) is inversely proportional to $D_{anchor}$ and decreases obviously along with the increase of $D_{anchor}$ as shown in Figure 7. Because PCFL (or ACFL) can convert three-dimensional localization algorithm to two-dimensional scenario, it can give accurate location and good reliability.

**The estimation and originality location**: In the simulation space, 45 mobile nodes are randomly distributed to localize themselves, the distances between each pair of them are 120m and 160m. As we can see in Figure 8, black rounds represent the original position, red and blue signs are the estimated coordinates computed by ACFL and PCFL, respectively, so it is quite clear that calculations of PCFL are more close to actual coordinates.

We can find that localization error of PCFL (or ACFL) reduces slightly when the number of samples is more than 400. When $D_{anchor}$ is 4 or larger, the localization error varies smoothly. Based on the running time of simulation experiments and localization error, we set the threshold for 0.0142, $D_{anchor}$ for 4, and the number of samples for 400 in the following simulation experiments.

**Figure 7.** The density of anchors

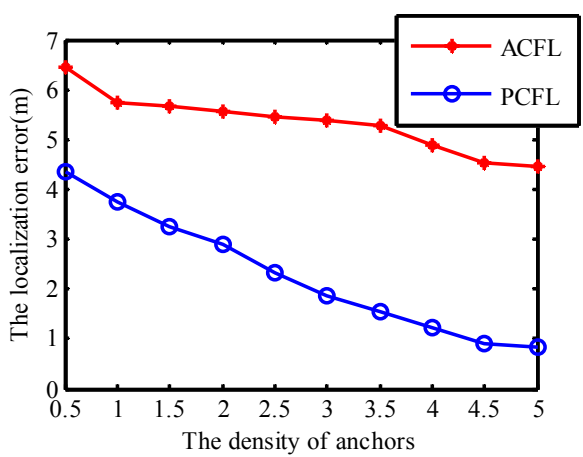

**Figure 8.** The originality and estimated coordinates

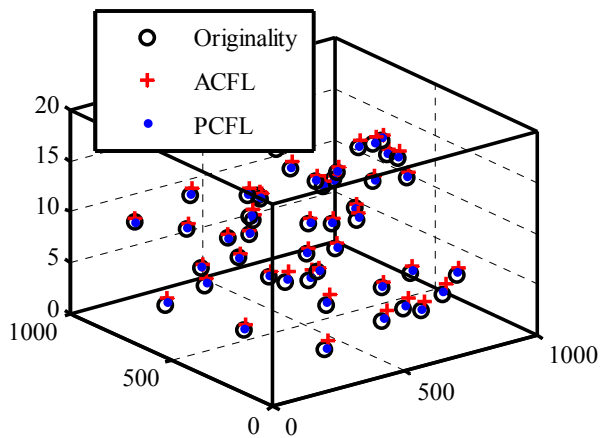

### 4.2. Comparison with different methods

Here, we run the simulation 100 times for each localization error in formula (29), then the average localization error can be obtained. We compare the five algorithms as shown in Table 2 which lists the maximum localization error, the minimum localization error, the average localization error and the standard deviation of the five algorithms.

The localization error of the Anchor-based algorithm is susceptible to the error of the estimated distances in relation to the coordinates of the fixed anchors, so the maximum localization error, the minimum localization error and the standard deviation of the Anchor-based algorithm are larger than the other four algorithms. Utilizing geometrical relationship, the AUV-aid method could perform localization coarsely, its average localization error is biggest. We can obtain that the average localization error of the ACFL is 4.56 which is larger than that of the AFLA and the PCFL by the simulation results. Table 2 shows that the PCFL algorithm has better localization performance and smaller localization error, comparing with the other four algorithms. The average localization error of PCFL can decline by about 30.4% than AFLA method.

**Table 2.** Comparison of localization errors

| Algorithm | Average error(m) | Max error(m) | Min error(m) | Standard deviation(m) |
|-----------|-----------------|--------------|--------------|----------------------|
| Anchor-based | 5.59 | 16.86 | 2.81 | 4.06 |
| AUV-aid | 9.82 | 7.72 | 1.56 | 2.69 |
| AFLA | 2.63 | 13.89 | 0.31 | 1.8 |
| ACFL | 4.56 | 10.22 | 0.51 | 2.06 |
| PCFL | 1.83 | 5.01 | 0.14 | 0.87 |

*4.3. The percentage distribution of the localization error*

Figure 9 shows the distribution histogram of the localization error, comparing among the Anchor-based algorithm, the AUV-aided algorithm, the AFLA algorithm, the ACFL algorithm and the PCFL algorithm. We deploy 20 mobile nodes stochastically in the simulation region, every mobile node performs self-localization simultaneously by running the independent experiment 50 times. Then the percentage distribution of the localization error of Anchor-based、AUV-aid、AFLA、ACFL and PCFL are shown in Figure 9. When the localization error of PCFL and the localization error of ACFL are both less than 2.5m, the percentage distribution is 88% and 23% in the process of simulation experiment, respectively. However the distribution percentage of the localization error of AFLA is less than that of PCFL. And the distribution percentage of the localization error of Anchor-based is also less than that of PCFL when the range of the localization error is from 0 to 2.5. The percentage distribution of the localization error of AFLA is the biggest compared with the other four algorithms when the localization error is varying from 2.5 to 5. The percentage distribution of PCFL is zero when the localization error is varying from 5 to 20. In a word, PCFL has better localization performance and smaller localization error.

**Figure 9.** The percentage distribution

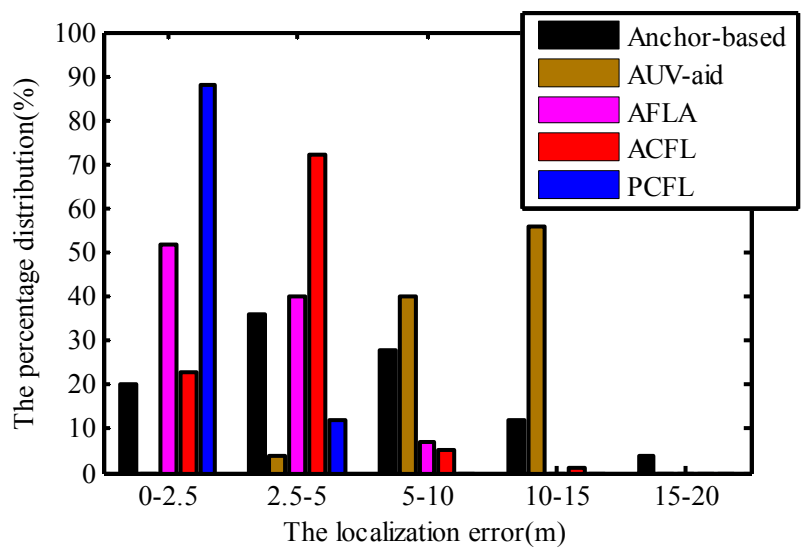

*4.4. Error and the speed of mobile nodes*

The speed of mobile nodes is an important factor affecting the localization error. Here the speed of mobile nodes is varying from 2m/s to 20m/s. The results of the simulation experiments indicate that the faster the moving speed, the bigger the localization error is. Due to using geometrical relationship to localize the mobile nodes coarsely, the localization error of AUV-aid is the maximum and changes stable without downtrend. AFLA used the geographical relationship of neighbor nodes to localize mobile nodes which did not need the information of anchors, so its precision is affected slightly by the speed of mobile node. Since the faster mobile nodes move, the less number of task anchors which can contact with them in three dimensional UASN can be acquired. So with the speed of mobile nodes increasing, the localization error of PCFL is almost equal to that of AFLA. And the overall varying trend of the localization error of PCFL or ACFL rises faster than the other three algorithms. When the speed is lower than 18 m/s, the localization error of PCFL is the minimum. As a whole, PCFL exceeds the other four methods with its average localization error of 3.92m, much smaller than that of ACFL, 7.33m, due to the projection technology.

**Figure 10.** The speed of mobile nodes

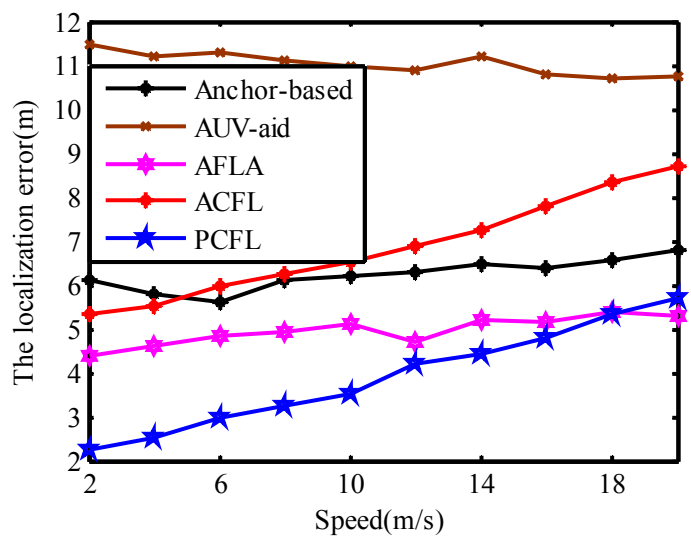

### 4.5. Error and the number of deployed mobile nodes

One hundred mobile nodes whose locations are unknown are randomly deployed in the three dimensional localization region. As we can see, in the process of simulation experiments, the five algorithms could localize all the deployed mobile nodes. Because the angles between task anchors and mobile nodes are calculated using the AOA method, the estimated distances between them have low accuracy and the localization error rises obviously with the number of the deployed mobile nodes increasing. The localization error of PCFL is smaller than that of the other four methods in Figure 11. The average localization error of AUV-aid is bigger than the other four methods. The error variation of the Anchor-based method fluctuates clearly and reaches the maximum value. Comparing with ACFL, PCFL is an effective self-localization algorithm, its localization error changes stable and keeps the smallest in the localization process.

**Figure 11.** The number of deployed mobile nodes

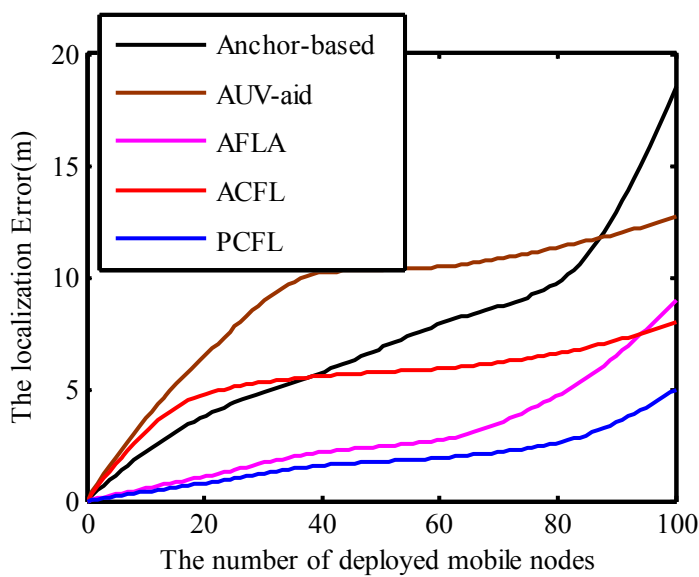

## 5. Conclusions

In this paper, two algorithms PCFL and ACFL using color filtering method have been proposed for mobile nodes self-localization in UASNs. The PCFL method is based on the RGB values of task anchors' projections, while ACFL is based on the RGB values of task anchors. The two methods can improve the coarse location by combining CDL method with the AOA values. Also the proportion factor of distance can optimize CDL method and the nearness degrees can help filtering samples more precisely. The proposed methods present better accuracy and robustness than Anchor-aid, AUV-aid and AFLA methods, especially PCFL method, when the speed of the mobile nodes is lower. In addition, the PCFL method is better than ACFL method. Furthermore, we plan to expand this work by the actual underwater experiments and reduce computational complexity of localization in the future.

**Acknowledgement**

This research was sponsored by the National Natural Science Foundation of China (Grant No. 61271125---"A Study of Mobile Node Localization Algorithms Based on Parameter Forecast and Sequence Filter for Wireless Sensor Networks"), Natural Science Foundation of Hebei Province (Grant No. F2013205084 ----"Research on Self-localization algorithms for Mobile Nodes in Wireless Sensor Networks") and the Educational Commission of Hebei Province (Grant No. Q2012124----"Research on the Localization for Mobile Sensor Networks Based on Sampling Filtering Theoretical Approach").

**Author Contributions**

All authors were involved in the mathematical developments and writing of the paper. The computer simulations were carried out by the Han Gao.

## Conflicts of Interest

The authors declare no conflict of interest.